\documentstyle[epsfig]{aipproc}
\begin{document}
 
\title{HEXTE Observations of SGR 1806--20 During Outburst}
 
\author{D. Marsden and R. E. Rothschild}
\address{Center for Astrophysics and Space Sciences, University of 
California at San Diego \\ La Jolla, CA 92093}

\author{C. Kouveliotou$^1$,S. Dieters$^2$}
\address{ES-84 NASA/MSFC \\ Huntsville, AL 35812}

\author{J. van Paradijs$^2$}
\address{Universitiet of Amsterdam, Astronomical Institute 
``Anton Pannekoek'' \\ Kruislaan 403, 1098 SJ, Amsterdam \\
\vspace{0.3in}
$^1$Universities Space Research Association \\
$^2$University of Alabama, Huntsville}

\maketitle

\begin{abstract}

We discuss observations of the soft gamma repeater SGR 
1806--20 during the RXTE Target of Opportunity observations made 
in November 1996. During the $\sim50$ ksec RXTE observation, 
HEXTE ($15-250$ keV) detected $17$ bursts from the source, with 
fluxes ranging from $3\times10^{-9}$ to 
$2.2\times10^{-7}$ ergs cm$^{-2}$ s$^{-1}$ ($20-100$ keV). We obtained 
spectra for the brighter HEXTE by fitting thermal bremsstrahlung and 
power law functions over the energy range $17-200$ keV. The best-fit 
temperatures and photon indices range from $30-55$ keV and $2.2-2.7$, 
respectively. The weighted average temperature of the HEXTE bursts 
was $41.8\pm1.7$ keV, which is consistent with previous SGR 1806--20 
burst spectra. The persistent emission from SGR 1806$-$20 was not 
detected with HEXTE.  

\end{abstract}

\section*{Introduction}

Soft gamma repeaters exhibit long periods of quiescence, often spanning 
years, punctuated by periods of intense bursting activity during which 
many brief (durations $<1$ s) and intense (luminosities L$\sim1-10^{3}$ 
L$_{Edd}$) bursts are emitted by the source \cite{norris91}. Believed 
to be neutron stars, the mechanism(s) for both the steady and bursting 
X--ray emission is still not well understood (\cite{thompson95}). 

SGR 1806-20 is the most prolific SGR, and it has been studied in 
the X--ray \cite{sonobe94}, optical \cite{vk95}, infrared 
\cite{kulkarni95}, and radio \cite{kulkarni94} bands. The source became 
active again during the Fall of 1996, emitting many powerful bursts that 
were first detected with BATSE \cite{kouv96}. A target of opportunity 
observation by the {\it Rossi X--ray Timing Explorer} (RXTE) was initiated 
on November 5, 1996. The data analyzed here were taken during that $50$ 
ks observation, which spanned the time interval starting at 10:53:20 UT 
(11/5/96) and ending at 10:52:00 UT (11/6/96).

\section*{Instrumentation} 

The HEXTE instrument \cite{gruber96} aboard RXTE consists of two 
clusters of collimated NaI/CsI phoswich detectors with a total net 
area of $\sim1600$ cm$^2$ and an effective energy range of $\sim15-250$ 
keV. The SGR observations discussed here were taken with the HEXTE in 
the $16$ second rocking mode, in which one cluster is always on the 
source, with the other pointed off-source for background accumulation. 
The clusters then beam-switch every 16 seconds, in such a way that one 
 cluster is always locked on-source. 

\section*{Results}

The on-source HEXTE data were binned into a time series for $3$ energy 
bands  ($15-50$ keV, $50-100$ keV, and $100-200$ keV). The $15-50$ keV 
time series for each continuous data stretch was searched for bursts 
using a Bayesian burst search algorithm developed at UCSD. The 
algorithm calculates the probability of a given number of bursts 
in each data stretch by incorporating the information on the expected 
background flux in each time bin. The burst search yielded $17$ bursts, 
all of which correspond to bursts seen by the PCA. The burst times and 
durations are shown in Table 1, and the lightcurve of two of the brighter 
bursts ($5$ \& $6$) is shown in Figure 1. The durations of the bursts 
seen by HEXTE ranges from less than $0.1$ to $0.6$ seconds, and the 
weakest burst detected by HEXTE corresponds to a PCA count rate of 
$800$ counts s$^{-1}$.
\begin{figure}
\centerline{\epsfig{file=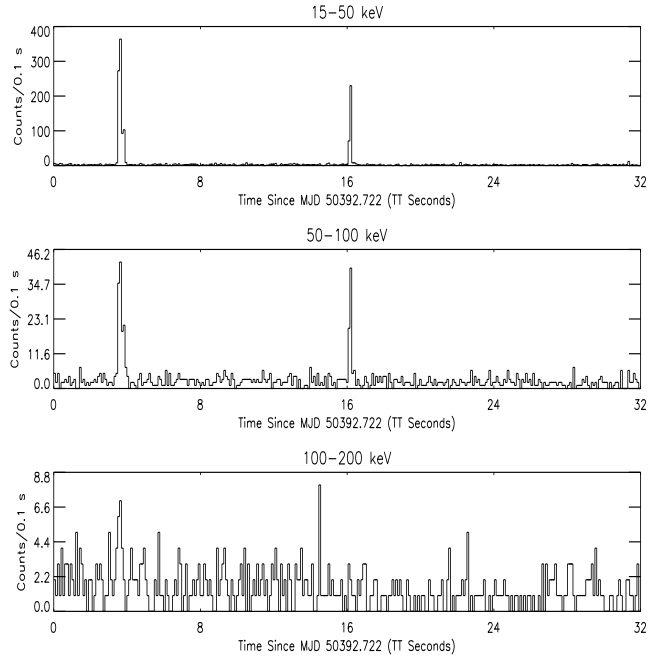,height=3.5in,width=3.5in}}
\vspace{10pt}
\caption{The lightcurves of SGR 1806--20 bursts $5$ \& $6$ as seen by 
HEXTE. The time resolution is $0.1$ s bins$^{-1}$.} 
\label{fig1}
\end{figure}

All of the HEXTE bursts were fit to thermal bremsstrahlung and power 
law functions using XSPEC, but only the $6$ brightest bursts 
yielded well-constrained spectral fits. In all the burst spectral fits, 
background was taken from {\it on-source} data taken immediately 
preceding and following the burst. The counts spectra were fit over the 
energy range $17-175$ keV, and the resulting best-fit parameters are 
shown in Table 1. The weakest SGR 1806--20 burst seen by HEXTE has a 
$20-100$ keV flux of $3\times10^{-9}$ ergs cm$^{-2}$ s$^{-1}$. 
Co-adding the 11 dim bursts and fitting them with power law 
and bremsstrahlung functions yields the best-fit parameters 
$\alpha=2.05\pm0.14$ and $kT=57\pm12$ keV for the mean spectrum of 
the weak HEXTE bursts. The HEXTE counts spectrum of a bright burst 
(Burst 2) is shown in Figure 2, with the best-fit thermal 
bremsstrahlung spectrum overplotted.
\begin{figure}
\centerline{\epsfig{file=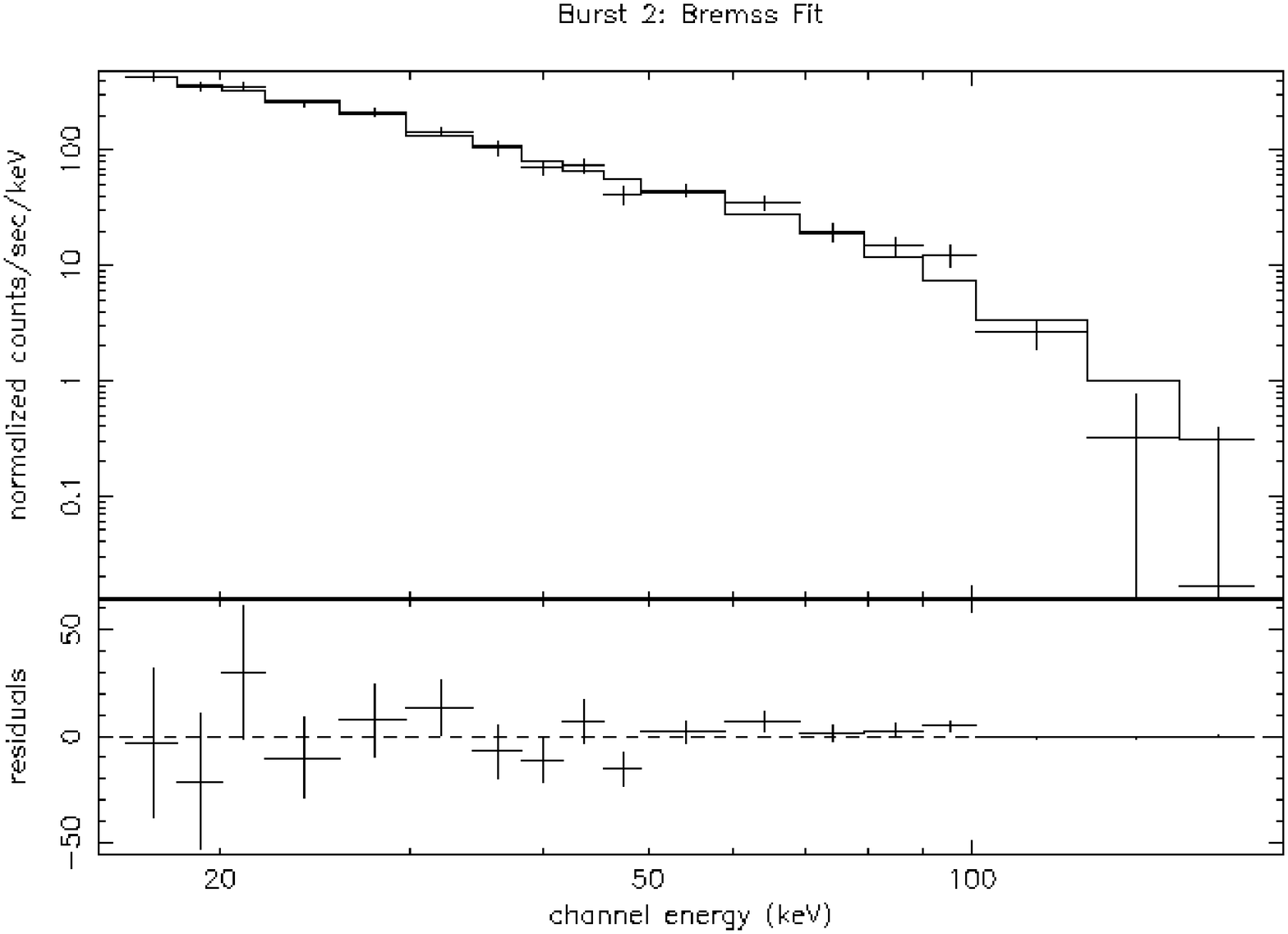,height=3.5in,width=4.in}}
\vspace{10pt}
\caption{The HEXTE count spectrum of a bright SGR 1806--20 burst 
(Burst 2). The solid line is the best-fit thermal bremsstrahlung 
model, and the residuals are shown in the bottom panel.} 
\label{fig2}
\end{figure}
The ${\chi}^{2}$ values were typically large for the stronger bursts, 
suggesting that more complicated spectral models may be needed to 
adequately fit the burst spectra.
\begin{table}
\caption{SGR 1806--20 bright HEXTE bursts}
\label{table1}
\begin{tabular}{ccddd} 
\multicolumn{1}{c}{Burst Time\tablenote{Terrestrial dynamical time 
(MJD modulo $50392$)}}& \multicolumn{1}{c}{Duration\tablenote{Seconds}}& 
\multicolumn{1}{c}{Photon Index\tablenote{Power law fit}}&  
\multicolumn{1}{c}{$kT$\tablenote{Thermal bremsstrahlung fit (keV)}}& 
\multicolumn{1}{c}{Flux\tablenote{$20-100$ keV flux ($10^{-8}$ 
ergs cm$^{-2}$)}}\\
\tableline
0.7106674 & 0.5 & 2.29$\pm$0.05 & 51.46$\pm$3.30 & 21.65$\pm$0.47 \\
0.7220713 & 0.5 & 2.52$\pm$0.07 & 37.62$\pm$2.76 & 12.53$\pm$0.68 \\
0.7222172 & 0.3 & 2.54$\pm$0.12 & 40.88$\pm$4.93 & 8.9$\pm$0.6 \\
0.7306593 & 0.2 & 2.16$\pm$0.19 & 54.19$\pm$14.47 & 5.1$\pm$0.5 \\
0.8404614 & 0.4 & 2.65$\pm$0.15 & 31.06$\pm$4.50 & 8.7$\pm$1.0 \\
1.1905632 & 0.3 & 2.32$\pm$0.14 & 45.67$\pm$7.85 & 10.86$\pm$0.11 \\
\end{tabular}
\end{table}

The weighted mean effective temperature of the six bright bursts and 
the fit to the mean spectrum of the weak HEXTE bursts is $kT=41.8
\pm1.7$ keV. This is consistent with previous measurements of SGR 1806--20 
burst temperatures \cite{atteia87}. A chi-squared test for a constant 
temperature yields ${{\chi_{\nu}}}^2=3.2$ for $\nu=6$, or a $0.5\%$ 
chance that the bursts all had the same temperature, suggesting that 
there may be some intrinsic variability in the burst spectra. 

\section*{Discussion}

The results of the HEXTE observations of SGR 1806-20 bursts are in 
general agreement with the durations \cite{kouv95} and bremsstrahlung temperatures \cite{atteia87} obtained by previous observers. The X--ray  luminosities ($20-100$ keV) of the bursts, in units of the Eddington 
luminosity, span the range $L_{X}\sim1-50$, assuming a distance to the 
source of $14.5$ kpc \cite{corbel97}, isotropic emission, and an 
Eddington luminosity of $10^{38}$ egs s$^{-1}$.

In the future, we plan on using the data from the Proportional Counter 
Array (PCA) aboard RXTE, in conjunction with the HEXTE data, to fit 
the SGR 1806--20 burst spectra over the entire $2-250$ keV energy range 
of the two instruments. This will result in a more accurate determination 
of the continuum spectral shape of the SGR bursts.
   
\paragraph*{Acknowledgements.} We thank NASA for support under grants
NAS5-30720 (D.M. and R.E.R.), NAG5-2560 (S.D. and C.K.), and 
NAG5-4878 (JvP)

\end{document}